\documentstyle[epsfig]{l-aa}

\begin{document}

   \thesaurus{(08.06.2;  % Stars: formation,  
               09.19.2;  % ISM: bubbles,
               09.19.2;  % ISM: supernova remnants,
               11.09.1 LMC;  % Galaxies: individual: LMC,
               11.19.4;  % Galaxies: star clusters,
               13.07.1)} % Gamma rays: bursts.
   \title{Gamma Ray Bursts versus OB Associations:
   do they trigger star formation?}

   \author{Yu. N. Efremov \inst{1},
           S.$\ $Ehlerov\' a \inst{2}, J.$\ $Palou\v s \inst{2}
           }
   \offprints{J. Palou\v s \inst{2} }
   \institute{\vskip3mm
\inst{1} Sternberg Astronomical Institute, MSU, Moscow 119899, Russia \\
\inst{2} Astronomical Institute, Academy of Sciences of the Czech Republic,
Bo\v cn\' \i \ II 1401, 141 31 Prague 4, Czech Republic \\
}

   \date{Received 11 March 1999/ Accepted 13 July 1999}

%   \markboth{Efremov et al.}{Gamma ray bursts vs. OB associations}

   \maketitle

   \markboth{Efremov et al.: Gamma ray bursts vs. OB associations}{}

\begin{abstract}
We discuss differences in shapes, expansion velocities and
fragmentation times of structures created by an energy deposition 
from a single
Gamma-Ray Burst (GRB) and an OB association to the ISM. 
After the initial inflation, supershells produced
by GRBs are almost static or slowly expanding, contrary to 
more rapidly expanding supershells created
by OB associations.
We discuss the position of the energy source
relative to the symmetry plane of the galaxy: observed arc-like structures
can be the most dense part of structures formed by an expansion
from a source above or below the galactic plane. Arcs may also form 
if the expansion takes place inside a giant HI cloud. We try to reproduce
the size, the age, and the average distance between OB associations in
the Sextant region at the edge of LMC 4. 

{\bf Stars: formation -- ISM: bubbles -- ISM: supernova remnants --
Galaxies: individual: LMC -- Galaxies: star clusters -- Gamma rays: bursts}

\vskip-7mm

   \end{abstract}

\section{Introduction} % 1

It is generally assumed that
the HI shells and supershells observed in galaxies may be formed by 
the action of massive stars in OB associations or they may be created by
impacts of high velocity clouds (HVC) into the galactic HI
plane. Recently, another possibility has been proposed: they may
result from an energy input
connected with Gamma-Ray Burst (GRB) events. This may
follow coalescence of two neutron stars in a binary system leading 
to the formation of a black hole.
Such events, lasting a few milliseconds, release up to 
$10^{54}$ erg of orbital kinetic energy of the progenitor. 
The energy may be released into a collimated jet
illuminating only a fraction $f_b$ of its sky.
Such beaming would decrease the output energy by a factor $f_b$ and
increase the frequency of occurrence by a factor $f_b^{-1}$.
Another way to release several times $10^{52}$ erg of energy is the 
collapse of a single super-massive ($\ge 100 M_{\odot }$) star, 
the hypernova (Paczynski 1998).   
X-ray afterglows, discovered by the satellite
BeppoSAX, lasting several hours, result from the
interaction  between relativistic particles and the ambient
medium around a black hole. The discovery of the host galaxies
in the optical domain also points to an interaction between
an energetic ($10^{54}$ erg) event  and the interstellar medium
in galaxies (Kulkarni et al. 1998; Wijers 1998). 
Other evidence in favor of such
superexplosions is given by Wang (1999), who discovered gaseous 
X-ray emitting remnants of two hypernovae in M101.  
Some  cold HI shells may be  the long
lasting evidence of a short pulse of energy accompanying
GRBs, as it has been noted by Blinnikov \& Postnov (1998) and 
elaborated by Efremov et al. (1998) and Loeb \& Perna (1998).

The intention of this paper is to demonstrate the difference
between HI shells created in connection to GRBs and HI shells 
formed around OB associations.
Properties of   supershells created in connection with a GRB
are  determined by a single short-acting source of energy, which
may arise in any position in a galaxy (and even outside it):
the energy of $10^{51} - 10^{54}$ erg is released in a few seconds
from a very small volume of the accretion disk around the black hole.
The activity of young stars in  OB associations is much less
concentrated in space and time: the initial volume can have a radius  
 $\ge 10$ pc and the energy may be delivered during several Myr.
The duration of the energy supply depends in this case on the history of the 
star
formation in the OB association and on the IMF, leading to a 
typical spread of 10 - 20 Myr.

In this paper we compare  simulations
of expanding shells connected with  OB associations to those created by
GRBs. We focus on the star
formation around the LMC4 region in the Large Magellanic Cloud (LMC). 
Next we discuss arguments for and against the origin of multiple
supershells from single superexplosion; and we consider the possibility 
that GRBs support star formation in galaxies.

\section{Simulations of shells created by a GRB and by an OB association} %2

The expansion of gas layers related to energy inputs from GRBs or
OB associations can be described as a blastwave propagating into the
interstellar medium of the host galaxy
(Ostriker \& McKee 1988; Bisnovatyi-Kogan \&
Silich 1995). Since the radius of
the shell is much larger than its thickness, the thin layer
approximation developed by Sedov (1959) can be used.
It
has been applied in one-dimensional models by Chevalier (1974) and in
two-dimensional models by Tenorio-Tagle \& Palou\v s (1987) and
Mac Low \& McCray (1988). These models have been further extended
into three dimensions by Palou\v s (1990) and Silich et al.
(1996).

A system of equations of motion,  mass, and  energy  is
solved.
The equation of motion
is
\begin{eqnarray}
{d \over dt} (m_{sh} v_{sh}) & =  & dS_{sh}[(P_{int} - P_{ext}) +
\rho _{ext} v_{ext} (v_{exp} - v_{ext})] \nonumber \\
& & + m_{sh} g,
\label{momentum}
\end{eqnarray}
where $m_{sh},\ v_{exp},\ dS_{sh}$ are the mass, the expansion velocity and
the surface of an element of the shell, 
$P_{int}$ and $P_{ext}$ are pressures inside and
outside the bubble, $\rho _{ext}$ and $v_{ext}$ are
the density and velocity of the ambient
medium, and $g$ is the gravitational acceleration.

The equation of mass conservation is
\begin{equation}
{d \over dt} m_{sh} = (v_{exp} - v_{ext})_{\perp } \rho _{ext}
dS_{sh} - m_{sh} \Gamma_{sh}.
\label{mass}
\end{equation}
The first term on the right hand side (rhs), giving the increase of mass
$m_{sh}$, is used  as long as the normal component of the velocity,
$(v_{exp} - v_{ext})_{\perp }$,
exceeds  the speed of sound in the ambient medium.
After the expansion becomes subsonic, the mass accumulation
stops. The second term on the rhs describes the decrease of
mass in the shell due to mixing with the material inside the
bubble; $\Gamma_{sh}$ is the shell mixing rate.

The total energy $E_{tot}$ is
\begin{equation}
E_{tot} = E_{pot}+E_{th}+E_{kin},
\end{equation}
where $E_{pot}$ is the gravitational potential energy in the galaxy. 
$E_{th}$ and $E_{kin}$ are the thermal and kinetic energies:
\begin{equation}
E_{th} = E_{th, int} + E_{th, sh} + E_{th, ext},
\end{equation}
\begin{equation}
E_{kin} = E_{kin, int} + E_{kin, sh} + E_{kin, ext},
\end{equation}
where the second parts of the subscripts on the rhs refer respectively to the
medium inside the bubble, in the shell, and in the unperturbed
environment. In this paper, we disregard the thermal energy of the
unperturbed ambient medium $E_{th, ext}$ and the kinetic energy
of the medium inside and outside the bubble $E_{kin, int}$
and $E_{kin, ext}$.

The balance of the internal thermal energy is given as
\begin{equation}
{d \over dt} E_{th, int} = L - \Lambda_{int} + \Gamma_{sh} E_{th, sh},
\label{then1}
\end{equation}
where $L$ is the energy input rate from a GRB or an OB association
and $\Lambda_{int}(\rho_{int}, T_{int}) $ is the radiative cooling rate
from the bubble interior, $\rho_{int}$ is the internal volume density,
$T_{int}$ is the internal temperature.
For the mixing rate we assume
\begin{equation}
\Gamma_{sh} = {f_{mix} \over t_{mix}},
\end{equation}
where
\begin{eqnarray}
f_{mix} = \cases{
1 & $T_{int} < T_{sh}$ \cr
0 & $T_{int} \ge T_{sh}$ \cr
},
\end{eqnarray}
$T_{sh}$ is the temperature in the shell.
The mixing time $t_{mix}$ is estimated as
\begin{equation}
t_{mix} = {\alpha r_{sh} \over V_{sound, sh}},
\end{equation}
where $r_{sh}$ is the radius of the shell and $V_{sound, sh}$ 
is the sound speed inside the shell:
$V_{sound, sh} = \sqrt{k T_{sh} \over \mu }$,
$k$ is the Boltzmann constant and $\mu $ is the mean atomic
weight of particles in the shell. $\alpha $ is a free parameter
giving the effectivity of the mixing process. $\alpha $ 
lies in the interval $(0, \infty)$:
\begin{itemize}
\item $\alpha = 0$ corresponds to an
instantaneous mixing, all the mass collected in the shell is
immediately mixed into the bubble
\item $\alpha = \infty $ corresponds to no mixing.
\end{itemize}
In our simulations we use $\alpha$ in the range (0.1 - 1), however results
are almost independent on the particular value, and moreover, mixing has 
influence only on the early evolution of the supershell. 
It does not have a substantial influence on the later evolution after 1 Myr
of expansion.

%%%%%%%%%%%%%%%%%%%%
%  fig 1
%%%%%%%%%%%%%%%%%%%%
\begin{figure*}[tbp]
\vglue-2cm
      \epsfig{figure=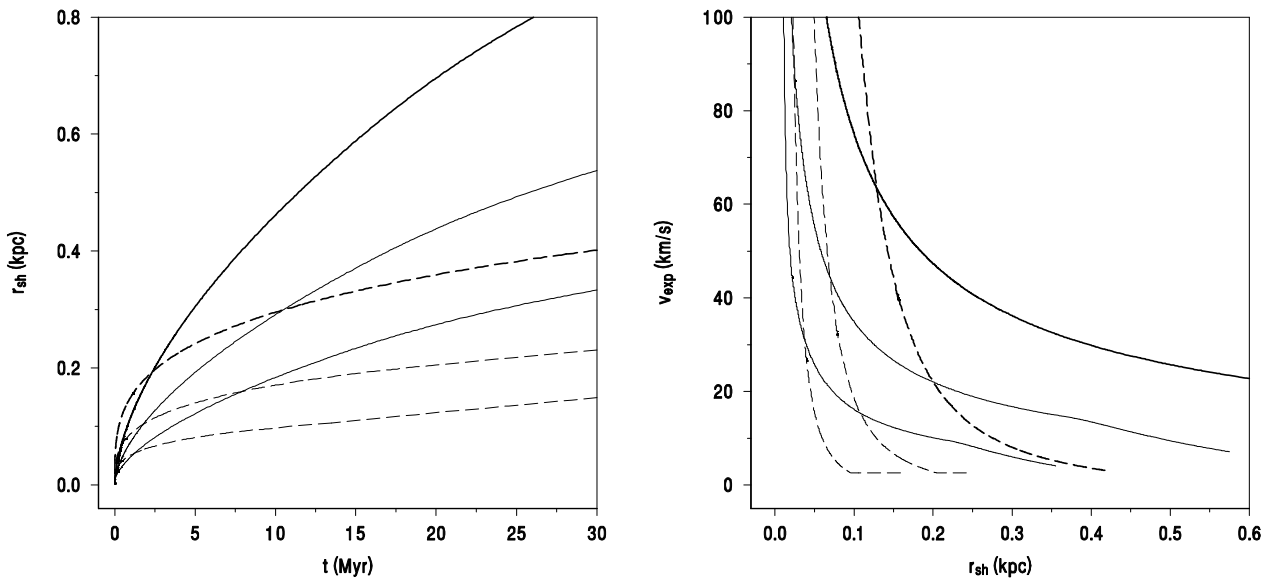,angle=270,width=18cm}%,height=6.5cm}
%      \picplace{0.01cm}   % standard 5.5cm
\vskip1cm
      \caption[]{The shell radius $r_{sh}$ as a function of time (left panel) 
and the
expansion velocity of the shell $v_{exp}$ as a function of radius 
(right panel) for shells expanding in a homogeneous medium with $\rho _{ext}$ =
1.4 cm$^{-3}$.
Solid lines denote the continuous energy input, the dashed ones
the abrupt energy input. The thickness of the line gives the input 
energy, from thin to thick:  $10^{52}, 10^{53}$ and $10^{54}$ erg.
}
      \label{fig1}
   \end{figure*}

The balance equation of the thermal energy in the shell is
\begin{equation}
{d \over dt} E_{th, sh} = {1 \over 2} {d m_{sh} \over dt} (v_{exp} -
v_{ext})_{\perp }^2 -\Lambda_{sh} - \Gamma_{sh} E_{th, sh},
\label{then2}
\end{equation}
where the first term on the rhs gives the fraction of the kinetic
energy of the shell which is converted into the shell thermal
energy due to the compression of the ambient medium.
$\Lambda_{sh}(\rho_{sh}, T_{sh})$ is the cooling rate of the
shell, where $\rho_{sh}$ is the volume density in the shell,
which is estimated under the assumption
that the thickness of the shell is a small fraction,
($\sim  0.1$), of the shell radius.
$T_{int}$ and $T_{sh}$ are derived from the thermal energies and number
of particles in the bubble $N_{int}$ and in the shell  $N_{sh}$:
$T_{int} = {E_{th, int} \over {3 \over 2} k N_{int}}$ and
$T_{sh} = {E_{th, sh} \over {3 \over 2} k N_{sh}}$.
The radiative cooling rate from both the bubble and the shell is
calculated using the cooling functions
of B\" ohringer \& Hensler (1989) for $T \in (10^{3.5} - 10^8)$ K,   
and Schmutzler \& Tscharnuter (1993) for $T \in (10^8 - 10^9)$ K.

The internal pressure  $P_{int}$ is derived from the internal thermal
energy $E_{th, int}$ and the volume of the bubble $V_{int}$
\begin{equation}
P_{int} = {2 \over 3} {E_{th, int} \over V_{int}}.
\label{pint}
\end{equation}

%The initial energy $E_0$ released from a GRB or from an OB association is
%divided among the internal $E_{th, int, ini}$ and the shell 
%$E_{th, sh, ini}$ initial thermal energies, and the initial 
%kinetic energy of the shell $E_{kin, sh, ini}$. 
%The initial temperatures are set to
%$T_{int, ini} = T_{sh, ini} = 10^9$ K. We assume:
%$E_{th, int, ini} = E_{th, sh, ini} = 0.35 E_0$, $E_{kin, sh, ini} = 
%0.3 E_0$.
%The mass from the initial  
%volume $m_{ini}$ within $r_{sh, ini}$ is divided equally between the 
%internal volume and the 
%shell: $m_{ini} = {4 \over 3} \pi r_{sh, ini}^3 \rho _{ext};  m_{int, ini} = 
%m_{sh, ini} = 0.5 m_{ini}$. With  the initial mass in the shell 
%$m_{sh, ini} = {2 \over 3} \pi r_{sh, ini}^3 \rho _{ext}$
%we derive the initial expansion velocity $v_{sh, ini}$ from
%$E_{kin, sh, ini} = 0.3 E_0 = {1 \over 2} m_{sh, ini} v_{sh, ini}^2$.
%The initial radius $r_{sh, ini}$ is derived so that the internal 
%initial mass 
%$m_{int, ini}$ can accommodate the initial internal thermal energy 
%$E_{th, int, ini}$ at the given initial internal temperature 
%$T_{int, ini}$:
%${2 \over 3} \pi r_{sh, ini}^3 \rho _{ext} \times {3 \over 2} 
%k T_{int, ini} = 0.35 E_0 $.

The distribution of the initial energy among the thermal and kinetic parts 
corresponds to the solution calculated by Sedov (1959), 
and the division of the 
initial thermal energy and mass between the internal volume and the shell is
quite randomly chosen. However, different ratios of initial masses 
and energies do not influence results substantially.

The shell is split
 into elements in a 3D space, and
equations of motion (\ref{momentum}), mass (\ref{mass}),
internal and shell thermal energy balance
(\ref{then1}) and (\ref{then2}), and the equation for the internal pressure
(\ref{pint}) are solved numerically  using finite timesteps.
To keep the accuracy of the integrated quantities under desired limits,
an adaptive step-size control scheme is used.

The expansion in the static, homogeneous medium with 
$\rho_{ext} = 1.4$ cm$^{-3}$ is shown in Fig.
\ref{fig1}. We compare results of the
abrupt and continuous energy input of $10^{52}$, $10^{53}$, and
$10^{54}$ erg:
\begin{itemize}
\item In the case of abrupt energy input, the
shell rapidly expands to the radius of 50 - 150 pc during the 
first $10^6$ yr. 
After 2 - 3 Myr the temperature $T_{int}$ decreases below $10^6$ K.
The initial inflation is followed by a slower
expansion at velocities $v_{exp}$ close to the speed of sound in 
the ambient medium.
\item
In the case of continuous energy input, the growth of the radius is
slower, reaching the size of 50 - 150 pc only after 2 - 3 Myr.
There is rapid growth phase as in the previous case, however, 
the expansion velocity
does not decrease as quickly. Already after 1 - 2 Myr it is higher
than in the case of abrupt energy input.
\end{itemize}
After $\sim$ 30 Myr, shells reach diameters of 100 - 350
pc and 300 - 1000 pc and masses of $10^{5.2} - 10^{7.1}$ and $10^{6.9} -
10^{8.1} M_{\odot}$  for, respectively, abrupt and continuous energy inputs 
of $10^{52} - 10^{54}$ erg. As seen in 
Fig. \ref{fig1}, shells connected to GRBs are   
smaller than shells connected to OB associations for   most of the time. 
At diameters larger than 500 pc, GRB shells always  
expand at  velocities  below 10 km s$^{-1}$. Thus,
large shells with
diameters 500 - 2000 pc expanding at velocities 10 - 40 km s$^{-1}$  
can not be created by an abrupt energy input connected to GRBs; they have to
be created by a continuous energy input related to massive stars in 
OB associations.  

%%%%%%%%%%%%%%%%%%%%
%  fig 2
%%%%%%%%%%%%%%%%%%%%
\begin{figure*}[hbtp]
\vglue-1cm
      \epsfig{figure=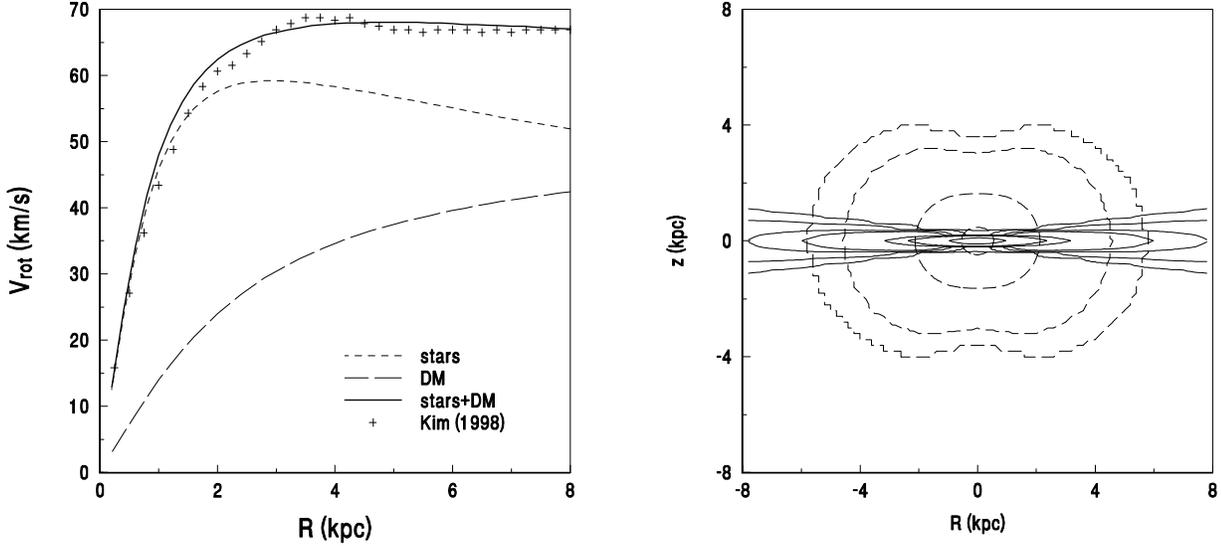,angle=270,width=18cm}%,height=6.5cm}
%      \picplace{0.01cm}   % standard 5.5cm
\vskip-1cm
      \caption[]{Left panel: a comparison of the observed rotation curve with 
a model of the LMC composed of dark matter and stars. 
Right panel:  the density
distribution in a two component disk
composed of cold gas - Solid line: (the values of the constant density are:
5.0, 1.0, 0.5, 0.1, 0.05, 0.01, 0.001 cm$^{-3}$) - supported almost 
completely by the
centrifugal forces,  and hot gas - Dashed line: (the values of the 
constant density are: 0.1, 0.01, 0.001, 0.0005 cm$^{-3}$) - balanced almost
completely by pressure gradients. }
      \label{fig2}
   \end{figure*}

\section{Expanding shells in the LMC}

In this section we apply the code described above  to expanding shells
in the Large Magellanic Cloud (LMC).
For the LMC we propose a three component model composed  of the interstellar
medium, stars and dark matter. We assume that the large-scale
shape of the gravitational potential is formed by stars and dark
matter: the contribution of the ISM  is neglected. 
The ISM contributes to local perturbations connected to instabilities and 
star formation.

\subsection{Stars and dark matter}

The spherically symmetric
volume density distribution of stars (subscript {\it S}) is
described by the Hubble-Reynolds profile
\begin{equation}
  \rho_S (R)  = {\rho_{S, 0} \over \left[ 1+\left( R
\over R_S \right) ^2 \right] ^{3/2}},
\end{equation}
and the less steep distribution of the dark matter (subscript
{\it DM}) with a function
\begin{equation}
  \rho_{DM} (R) = {\rho_{DM, 0} \over 1+\left( {R
\over R_{DM} } \right) ^2 },
\end{equation}
where $\rho $ is the volume mass density, $R$ is the
galactocentric distance. $R_S$ is the core radius of the stellar
distribution and $R_{DM}$ the radial scale length of the
dark matter. The subscript $0$ denotes values at
the galactic center.

The stellar  + dark mass within the radius $R$ is
\begin{eqnarray}
%\begin{flushleft}
 & M_{S+DM}(R)  =  4 \pi R^3_S \rho _{S, 0} \times \nonumber \\
 & \left ( ln[{R \over R_S} + ({R^2
\over R_S^2} + 1)^{0.5}]
- {R \over R_S} [({R \over R_S})^2 + 1]^{-0.5}
\right ) \nonumber \\
  & +  4 \pi R^3_{DM} \rho _{DM, 0}
\left ( { R \over
R_{DM} } - arctan ( {R \over R_{DM} } ) \right ).
%\end{flushleft}
\end{eqnarray}

The galaxy extends up to the galactocentric distance $R_B$:
\begin{equation}
\rho = \rho _S + \rho _{DM} = 0 \ \ \ {\rm for} \ \ \    R > R_B.
\end{equation}

To balance the radial force  a circular velocity $V_{rot} (R)$ is
required:
\begin{equation}
V_{rot}(R) =\left ( {G M_{S+DM} (R) \over R}\right ) ^{1/2},
\end{equation}
where $G$ is the gravitational constant.

The potential is given as
\begin{equation}
\phi = - {V_{rot} (R_B)^2 \over 2} F(R),
\end{equation}
where
\begin{eqnarray}
  F(R) =   1  +
  {1 \over 2}{R_B \over R_{DM}} {M_{DM} \over (M_{DM} + M_S)}
\times & \nonumber  \\
  \left [ ln(1 + \left ( {R_B \over R_{DM}}
\right )^2 ) - ln ( 1 + \left ( {R \over R_{DM}} \right )^2 ) \right ]
& \nonumber \\
  +  {R_B \over R_{DM}} {M_{DM} \over (M_{DM} + M_S)}
\left [ {arctan ( {R_B \over R_{DM}}) \over {R_B \over R_{DM}}}
- { arctan ({R \over R_{DM}}) \over {R \over R_{DM}}} \right ]
& \nonumber \\
  +  {R_B \over R_S}{M_S \over (M_S + M_{DM})}
\times & \nonumber \\
 \left [ {ln \left (
{R \over R_S} + \sqrt{ 1 + \left ( {R \over R_S} \right )^2 }
\right )
\over {R \over R_S}}
- {ln \left (
{R_B \over R_S} + \sqrt{ 1 + \left ( {R_B \over R_S} \right )^2 }
\right ) \over {R_B \over R_S}} \right ] &  \nonumber \\
 \ \ \ {\rm for } R \leq R_B \ \ \ & \nonumber   \\
 F(R) =  {R_B \over R}  \ \ \ {\rm for} \ \ \ R > R_B.\ \ \ \ 
 &
\end{eqnarray}

The rotation curve of the LMC given by Kim et al. (1998) is
reproduced with $R_S = 1$ kpc, $R_{DM} = 2$ kpc, $R_B
 = 8$ kpc,
and the total mass $M_{S+DM} (8\ {\rm  kpc}) = 8 \times  10^9
M_{\odot }$. $60 \%  $ of this mass is in stars and $40 \% $ in dark
matter (see Fig. \ref{fig2}).

\subsection{Interstellar medium}

The interstellar medium is distributed in a two component disk,
which is partially
supported by the centrifugal force and partially balanced by the
pressure gradients:
\begin{equation}
V_G (r)  = \left ( e^2 r {\partial \phi \over \partial r} \right
) ^{1/2},
\end{equation}
where $V_G$ is the plane parallel rotational velocity of the gas
at a distance
$r$ from the rotational axis. A fraction $e^2$ of the
gravitational force ${\partial \phi \over \partial r}$ towards the
rotational axis is balanced by the centrifugal force, the
complementary
fraction $(1 - e^2)$ is balanced by
the pressure gradients. $e$ is a free parameter from the interval
$0 < e < 1$: $e = 1$ means that the gravity is completely
balanced by the centrifugal force,  $e = 0$ means that the
gravity is completely balanced by  pressure gradients.
This model has been  described by Tomisaka \& Ikeuchi
(1988) and used by Suchkov et al. (1994) and Silich \&
Tenorio-Tagle (1998) for models of large-scale bipolar flows from
nuclear starburst in dwarf galaxies.

The distribution of the gas is given as
\begin{equation}
\rho _G = \rho _{cold} +\rho _{hot},
\end{equation}
with
\begin{equation}
\rho _i = \rho _{i, 0} \  exp ({\it const}_i\  \chi ),
\end{equation}
where $i$ stands for $cold$ or $hot$ gaseous components.
${\it const}_i$ is inversely proportional to the square of the sound speed
$c_S$ and to the temperature $T$  of the gaseous component, and
\begin{equation}
\chi  = \phi (R) - e^2 \phi (r) - (1 - e^2) \phi (0).
\end{equation}
The cold and hot gaseous components vary  in the degree of 
pressure support:
\begin{itemize}
\item the cold component, where we assume $c_S = 8$ km s$^{-1}$
(corresponding to $T = 10^4$ K), is distributed in a disc
mainly supported by its rotation: $e = 0.9$;
\item the hot component, where $T = 10^6 $ K,
is more supported by pressure gradients: $e = 0.5$, resulting
in an almost spherical distribution.
\end{itemize}
Densities of both components in a model of the LMC are shown in
Fig. (\ref{fig2}).
With $\rho _{cold, 0} = 11$ cm$^{-3}$ and $\rho _{hot, 0} =
0.15$ cm$^{-3}$,
the total mass  of the cold and hot component within 4 kpc is
$5.2 \times 10^8 M_{\odot }$, respectively $5.2 \times 10^7 M_{\odot }$.

\subsection{Fragmentation of expanding shells}

Conditions for the instability and fragmentation of expanding
decelerating shells have been expressed by Ehlerov\' a et al.
(1997) using results of the linear perturbation theory of transverse motions
on a three-dimensional shell expanding into a uniform ambient
medium, as discussed by Elmegreen (1994) and Vishniac
(1994). The stretching of the perturbed region due to the expansion
may be compensated by the convergence of the flow due to the self-gravity
of the shell, if the maximum growth rate  of a
transverse perturbation $\omega > 0$.
\begin{equation}
\omega = -{3v_{exp} \over r_{sh}} + \sqrt{{v^2_{exp} \over r^2_{sh}} +
\left ({\pi G \Sigma
\over c}\right)^2},
\label{condx}
\end{equation}
where  $\Sigma $ is the mass column density of the shell, $c$
is the speed of sound within the shell.
The wavelength of the fastest transversal perturbation, $\lambda $,
is given  as
\begin{equation}
\lambda = { 2 c^2 \over G \Sigma }.
\label{lamb}
\end{equation}
$\omega $ and $\lambda $ can be evaluated in 3D
computer simulations using the thin shell approximation. In the
beginning of an expansion, when $v_{exp}$ is large and $r_{sh}$ is small,
$\omega $ is always negative, and the shell is stable. The
stretching because of the rapid expansion at early stages is 
much more important than the self-gravity of the shell.
Later, $v_{exp}$ decreases and $r_{sh}$ increases, so that the
first negative term in
Eq. (\ref{condx}) becomes less important, and at the same
time $\Sigma $ increases so that the second term in
Eq. (\ref{condx}), which is always
positive, becomes more important.
At the expansion time $t_b$, $\omega $ becomes positive  
and we start to 
evaluate the fragmentation integral $I_f(t)$:
\begin{equation}
I_f(t) = \int _{t_b}^{t} \omega (t')dt'.
\label{tfrag}
\end{equation}
At the time $t = t_f$ when $I_f(t) = 1$, fragments are already well
developed.
We also calculate the wavelength $\lambda $ at the times $t_b$ and
$t_f$.

\begin{table*}%[h]
 \centering
 \begin{tabular}{| c | c c c | c c c | c c c |}
 \hline
 $\rho_{ext}$ & type & $E_0$  & $r_{ini}$ & $t_b$ & $r(t_b)$ & 
$\lambda (t_b)$ &
$t_f$ & $r(t_f)$ & $\lambda (t_f)$ \\
 $cm^{-3}$ & & {\it erg} & {\it pc} & {\it Myr} & {\it pc} & 
{\it pc} & {\it Myr} & {\it pc} & {\it pc} \\ 
\hline
 1.4 & 
GRB & $10^{52}$ & 4 &  &  &  &  &  & \\
 & GRB & $10^{53}$ & 9 & & & & & & \\
 & GRB & $10^{54}$ & 20 & 8.8 & 252 & 99 & 32 & 632 & 14 \\ 
 & SN & $10^{52}$ & 2 &  17 & 207  & 121 & $ >35$ & & \\

 & SN & $10^{53}$ & 2 & 13 & 276 & 83 & 35 & 348 & 105 \\
 & SN & $10^{54}$ & 2 & 9.1 & 334 & 64 & 23 & 413 & 59 \\
\hline
10 & GRB & $10^{52}$ & 2 & & & & & & \\
 & GRB & $10^{53}$ & 5 & 5.4 & 86 & 48 & 26 & 128 & 86 \\
 & GRB & $10^{54}$ & 10 & 3.7 & 133 & 30 & 11& 175 & 22 \\
 & SN & $10^{52}$ & 1  & 8.9 & 110 & 37 & 18 & 159 & 24 \\
 & SN & $10^{53}$ & 1  & 6.6 & 145 & 27 & 13 & 210 & 17 \\
 & SN & $10^{54}$ & 1  & 4.7 & 186 & 20 & 9.5 & 264 & 13 \\
 \hline
20 & GRB & $10^{52}$ & 2 & & & & & & \\
 & GRB & $10^{53}$ & 4 & 3.7 & 67 & 31 & 15.5 & 100 & 44 \\
 & GRB & $10^{54}$ & 8 & 2.5 & 102 & 20 & 7.2 & 136 & 14 \\
 & SN  & $10^{52}$ & 1 & 6.5 & 80  & 26 & 13  & 119 & 17 \\
 & SN  & $10^{53}$ & 1 & 4.9 & 107 & 19 & 9.7 & 157 & 12 \\ 
 & SN  & $10^{54}$ & 1 & 3.5 & 138 & 14 & 7.0 & 202 & 9  \\
 \hline
30 & GRB & $10^{52}$ & 2 & & & & & & \\
 & GBR & $10^{53}$ & 3 & 3.0 & 57 & 25 & 10.0 & 79 & 27 \\
 & GRB & $10^{54}$ & 7 & 2.0 & 88 & 16 & 5.8  & 117 & 11 \\
 & SN  & $10^{52}$ & 1 & 5.4 & 67 & 21 & 11   &  99 & 14 \\
 & SN  & $10^{53}$ & 1 & 4.0 & 89 & 15 & 8.1  & 132 & 10 \\
 & SN & $10^{54}$  & 1 & 2.9 & 115 & 12 & 5.9 & 171 & 7 \\
 \hline
 \end{tabular}
 \caption{3D simulations of expanding shells  in the model of the
LMC. $\rho_{ext}$  is the density of the ambient medium 
at $R_0 = 2$ kpc, $z = 0$ kpc. The value $1.4\  cm^{-3}$ corresponds to the 
described model, other values are artificially increased.
GRB means an abrupt energy input, SN a continuous one.}
\label{fragtab}
\end{table*}

\subsection{Expanding shells}

%%%%%%%%%%%%%%%%%%%%
%  fig 3
%%%%%%%%%%%%%%%%%%%%
\begin{figure*}
\begin{center}
      \epsfig{figure=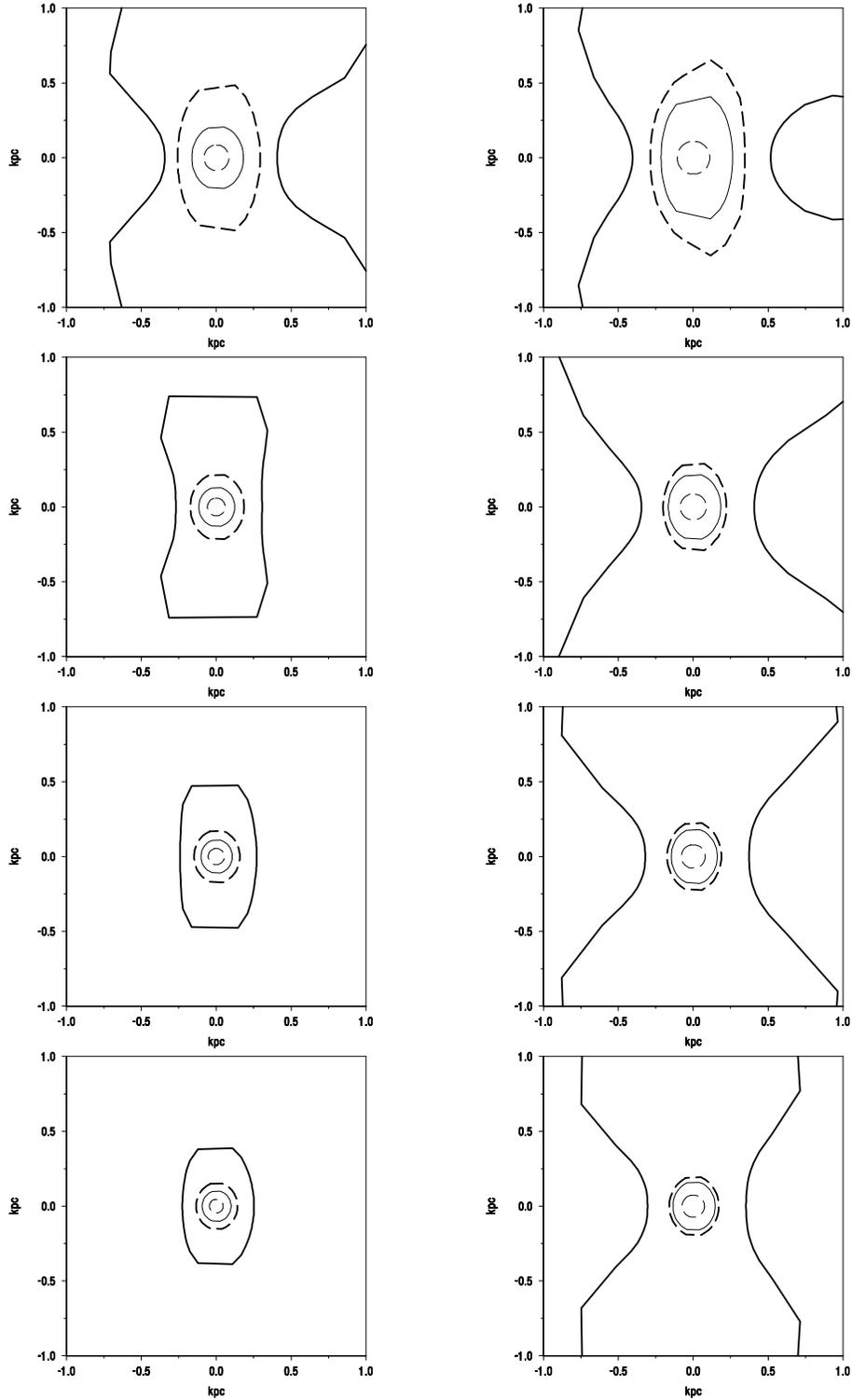,angle=180,width=12cm}%,height=6.5cm}
\vglue-2cm
%\hglue-2cm
%      \picplace{0.01cm}   % standard 5.5cm
\end{center}
\vskip1cm
      \caption[]{$R - z$ cuts of  expanding shells at a
distance of 2 kpc from the center of the LMC after
10 Myr (left panel) and 20 Myr (right panel) of evolution. 
The galactic center is in the direction to the left of the (0, 0) point.
The density of the
ambient medium $\rho_{ext}(z=0)$ from the upper to the bottom row is: 
1.4, 10, 20,
30 cm$^{-3}$. 
Thin and thick lines correspond to an input energy of $10^{52}$
and $10^{54}$
erg respectively, solid and dashed lines mean a continuous or abrupt 
energy input.}
      \label{fig3}
   \end{figure*}

We have done computer simulations listed in
Table \ref{fragtab}. 
The value of $\rho_{ext} = 1.4$ cm$^{-3}$ corresponds to the density at
a distance of 2 kpc from the center of the LMC (and $z = 0$). 
We calculated experiments with higher densities of the 
ambient medium, too. The density could be locally increased as a result of
sweeping of the mass from other places.   

Due to gradients of the ISM density
and to the form of the gravitational potential, shapes of expanding
shells are not spherical; they are usually elongated in the direction 
perpendicular to the galaxy plane.  This is
demonstrated in the $R - z$ cuts through expanding shells,
which are shown after
10 and 20 Myr of expansion in Fig. \ref{fig3}. For the continuous 
energy input only the low energy  cases, $E_0 = 10^{52}$ erg, do not blow-out 
to the galactic halo. For $E_0 \ge 10^{53}$ erg bubbles blow-out to high 
$z$ distances. With the low (not enhanced) value of the density 
of the ambient medium, the blow-out is radially asymmetrical, as can be 
seen in Fig. \ref{fig3}. 
This radial asymmetry is the result of gradients in the distribution of the 
hot 
gaseous component, which produce the outflow in the direction away from 
the galactic center, as seen at high $z$-distances in Fig. \ref{fig3}. 
This effect may result in a 
systematic difference between observed shapes of supershells at the near and 
far side of the LMC: they should be more round in the north-eastern part of 
the LMC than 
in the south-western one (for the discussion of the LMC spatial
orientation see Sect. 4). As an example the nearly circular supergiant 
shell SG24 in the NE  and the elliptical shell SG8 in the SW can be given 
(Kim 1998).  
The abrupt energy input creates much
less asymmetrical bubbles, which even for higher energies do not blow-out to
the galactic halo. These shells stay near the galaxy plane even for 
the highest input energies 
and lowest external densities; they never reach the height of 1 kpc from the 
$z = 0$ plane. This is a more general result, connected to the initial
rapid expansion and subsequent decrease of the shell expansion velocity 
for the abrupt
energy input. GRB can lead to a blowout if they explode below or above the 
symmetry plane of the galaxy, or
if they release high energies ($ \sim 10^{54}$ erg) in 
thin-disk spiral galaxies.

%%%%%%%%%%%%%%%%%%%%
%  fig 4
%%%%%%%%%%%%%%%%%%%%
\begin{figure*}
\vglue-2cm
      \epsfig{figure=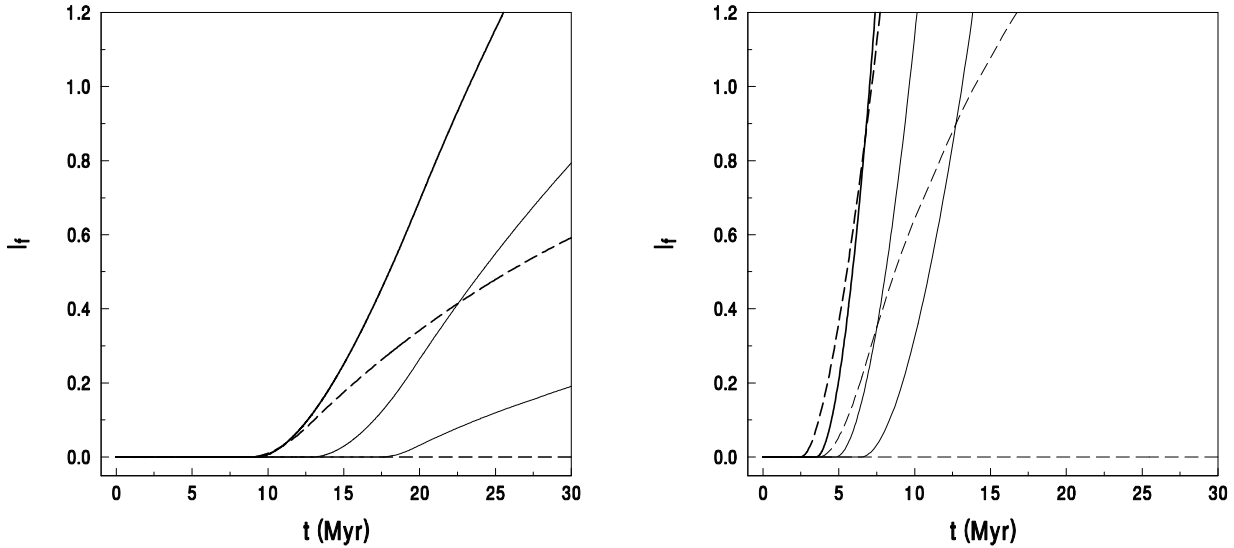,angle=270,width=18cm}%,height=6.5cm}
%      \picplace{0.01cm}   % standard 5.5cm
\vskip1cm
      \caption[]{The fragmentation integral $I_f$ as a function
of time for $\rho_{ext} = 1.4$ cm$^{-3}$ (left panel) and
$\rho_{ext} = 20$ cm$^{-3}$  (right panel). Designation of lines is the 
same as in Fig. \ref{fig1}.}
      \label{fig4}
   \end{figure*}

In Table
\ref{fragtab} we give fragmentation times $t_b$ and $t_f$ and 
wavelengths of the fastest perturbation, $\lambda $, given by Eq. 
(\ref{lamb}).
We add simulations with  low input energy,  
$E_0 = 10^{51}$ erg, corresponding to the energy of one supernova. 
For this input, energy shells are small and always stable for 
any tested density. It means that an isolated SN explosion
or GRB of the same energy can not induce gravitational 
fragmentation and trigger star formation. To do so in the LMC 
(and elsewhere) higher input energies are needed.

Shells related to continuous energy inputs can fragment for 
lower input energies than shells created by abrupt ones: in the first case 
$E_0 \ge 10^{52}$ erg is enough for a  density of the ambient medium in the 
range $\rho_{ext}$ = 1 - 30 cm$^{-3}$.
In the second case $E_0 \ge 10^{53}$ erg is needed for 
$\rho_{ext} \ge 10$ cm$^{-3}$, and $E_0 \ge 10^{54}$ for  
$\rho_{ext}$ = 1.4 cm$^{-3}$.   

The time evolution of the fragmentation integral $I_f$ is shown
in Fig. \ref{fig4}. In the low density ambient medium shells  
are stable for quite a long time. Except for the continuous input of $10^{54}$
erg,  neither of the cases reaches $t_f$ before 30 Myr, which means, that
the fragmentation is very slow if it happens at all.

For higher values of the density of the ambient medium, the 
shells fragment. For input energies $E_0 \ge 10^{53}$ erg, the instability 
starts earlier for abrupt energy input cases, 
but it proceeds faster in continuous cases than in their abrupt counterparts.

\section{Triggered star formation at the edge of LMC 4 in the
Sextant region}

In this section  we focus on a region in the LMC near 
Shapley's Constellation III, where several OB associations have
been identified by Lucke \& Hodge (1970). A supergiant HI shell
surrounding the most prominent HI void in the LMC can be seen in e. g.
Kim et al. (1998) at the same location.
At the SW edge of this shell, outside the HI void, OB associations 
LH51, LH54, LH60w, LH60e, and LH63 form a fraction of the arc. This arc 
has been  named the Sextant by Efremov \& Elmegreen (1998a), since the 
OB stars make about 1/6 of the complete circle. Its radius is 
estimated to be $\sim $170 pc and the average distance between subsequent OB 
associations in the arc is $\sim 32$ pc.     
Near the center of the Sextant, two star clusters (HS288 and HS287)  
have been found. The oldest stars in the Sextant are $\sim 7$ Myr old. 
We assume the age of the two star clusters to be 
10 - 15 Myr. The direct determination of ages of these clusters 
is desirable.  

Here, we would like 
to discuss once more the scenario proposed by Efremov \& Elmegreen (1998a), 
i.e. that $\sim $ 15 Myr ago the energy has been released from star 
clusters HS288 
and HS287 and formed an expanding shell. Later, the shell  fragmented 
and triggered the formation of OB associations, which during the following
 7 Myr moved to the presently observed configuration, 
the Sextant.  An alternative concept is the energy injection from 
a merging event between two compact components of the close binary system
related to a GRB: the progenitor was probably ejected from the 
nearby massive
star cluster NGC 1978 (Efremov 1998, 1999a; Efremov \& Elmegreen, 1998b) 
as a consequence of SN explosions leading to the formation of
compact components. 
If this is the case, HS287 and HS288 have nothing to do with the Sextant. 
It may explain why the two clusters are slightly off  center with respect to 
the  Sextant arc.
Some questions remain unanswered in both scenarios: 
particularly, why we see only a fraction 
of the circle, and how its orientation relative
to the line of nodes of the LMC (see below) may be explained.

We try to reproduce the shape of the Sextant, its radius, age and  
the distances between the OB associations using
numerical simulations of  
shells expanding in a model of the LMC, taking into account
different positions relative to the symmetry plane.
An alternative scenario is the supershell expanding in   
the preexisting HI supercloud, which may be a result of the gravitational 
fragmentation in walls of the HI ring surrounding the 
LMC 4 region. 
 
The position of the LMC on the sky is specified by the line of nodes 
$\theta _0$ and 
the inclination $i$. We adopt values given by Luks \& 
Rohlfs (1992): $\theta _0 = 162^{\circ }, i = 33^{\circ }$. 
An analysis of proper motions of LMC stars in the Hipparcos catalogue
revealed (Kroupa \& Bastian 1997), that the LMC rotates clockwise. In the 
combination with the analysis of radial velocities by Luks \& Rohlfs (1992) 
it means, that the SW half   is the more distant part of the LMC 
and the NE half is the nearby one. 

The Sextant itself is almost
perpendicular to the line of nodes with the radius parallel to it. 
It is also interesting, that a continuation of the Sextant arc 
would cross the 
LMC 4 HI ring nearly perpendicularly.    
Therefore, its  radius of 170 pc is almost uninfluenced by  
projection effects,
while the deprojected average distance between the OB associations in the 
Sextant is $l_{OB} \sim 37$ pc.

Using computer simulations, we search for such initial energies and  
densities of the ambient medium to get unstable shells with radii of 170 pc.
With a shell of radius 170 pc we ask what was the value of the fragmentation 
integral $I_f(t)$ some 7 Myr ago (7 Myr is the age of the oldest stars in the 
Sextant). We request a value between 0 and 1.
For small values of $I_f$ (near 0) the formation of stars must have been 
a very quick process, since they formed very shortly after the 
shell started to be unstable. For $I_f$ near  1 the star formation was 
somewhat delayed. If $I_f$ = 0 about 
7 Myr before reaching the radius 170 pc, shells do not fragment 
early enough and cannot form OB stars in 
the Sextant. If $I_f > 1$  7 Myr 
before reaching 170 pc, shells also cannot reproduce the Sextant, since
stars would probably form earlier and the observed arc of 
OB associations would be composed of older stars.   

We ask what is the ratio between the wavelength of the fastest transversal
perturbation in the shell, $\lambda $, and the shell radius $r_{sh}$ 7 Myr
before reaching the radius of 170 pc. In our opinion, this value did 
not change 
since the formation of the first stars. Therefore, it should reproduce 
the presently observed 
ratio of the average deprojected distance between the OB associations 
$l_{OB}$ to the radius of the Sextant: ${37 \over 170} = 0.22$.
   
The above conditions restrict the possible values of the input energy and 
of the density of the ambient medium:
\begin{itemize}
\item The continuous energy input forms shells of  a nearly correct size 
for $E_0 = 10^{53}$ erg and $\rho_{ext}$ = 20 cm$^{-3}$:  at \ t = 4 Myr 
$r_{sh}$ = 105 pc, $I_f$ = 0.1, and $\lambda / r_{sh}$ = 0.18. 
The radius of 170 pc is reached  11 Myr after the beginning of the expansion.
Other simulations do not meet the conditions so well.       
\item The abrupt energy input gives the best solution with $E_0 = 10^{54}$
erg and $\rho_{ext}$ = 10 cm$^{-3}$. At t = 3 Myr $r_{sh}$ = 130 pc, 
$I_f = 0.1$, and $\lambda / r_{sh} = 0.22$. The radius 170 pc is reached at
t = 10 Myr.  Lower energies or higher densities 
do not fulfill the above conditions. 
\end{itemize}
We conclude, that for both continuous and abrupt energy input we can 
reproduce the correct size, age and average distance between fragments, 
observed in the Sextant.
The continuous energy input gives the best solution for an input energy
of $10^{53}$ erg and density of the ambient medium 
20 cm$^{-3}$. In the case of abrupt energy input, a higher value of energy  
($10^{54}$ erg) and a lower density of the ambient medium 
($\sim 10$ cm$^{-3}$) are needed.
        
%% fig5
%%%%%%%%%%%%%%%%%%%%%%%%%%%%%%%%%%%%%%%%%%%%%%%%%%%%%%%%%%%%%%%%
\begin{figure}[htbp]
\leavevmode
%\vglue-1cm
\epsfig{file=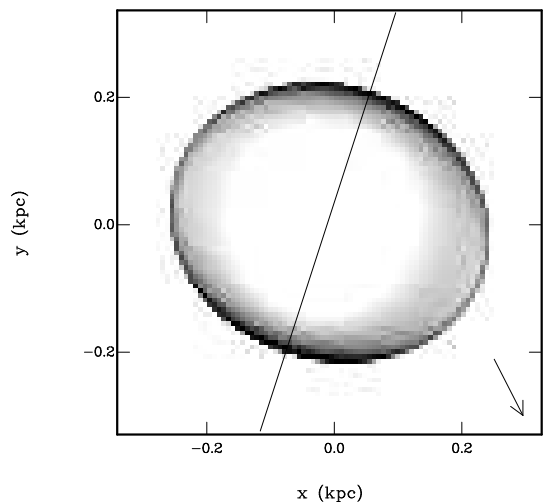, width=8cm, angle=0}
%\vskip-5cm
\caption[]{Projected column density of an expanding shell created by 
the continuous
energy input from the expansion center at the symmetry plane of the galaxy.
North is up and West is right, the line shows the direction of the 
line of nodes, the 100 pc long arrow points towards the 
center of the LMC.}
\label{sym}
\end{figure} 
%% fig6
%%%%%%%%%%%%%%%%%%%%%%%%%%%%%%%%%%%%%%%%%%%%%%%%%%%%%%%%%%%%%%%%%
\begin{figure*}[htbp]
\leavevmode
%\vglue-1cm
\epsfig{file=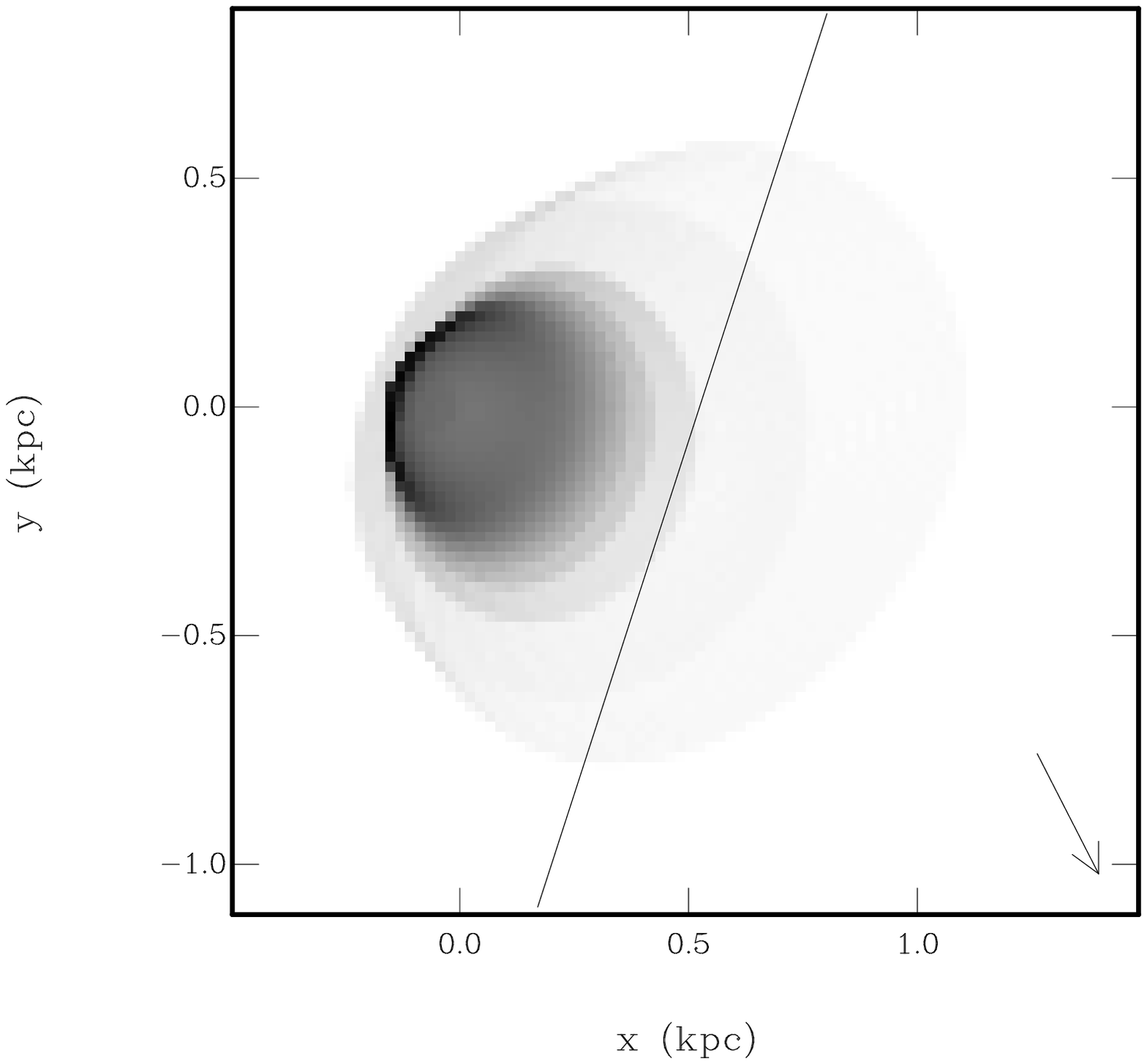, width=8cm, angle=0}
\epsfig{file=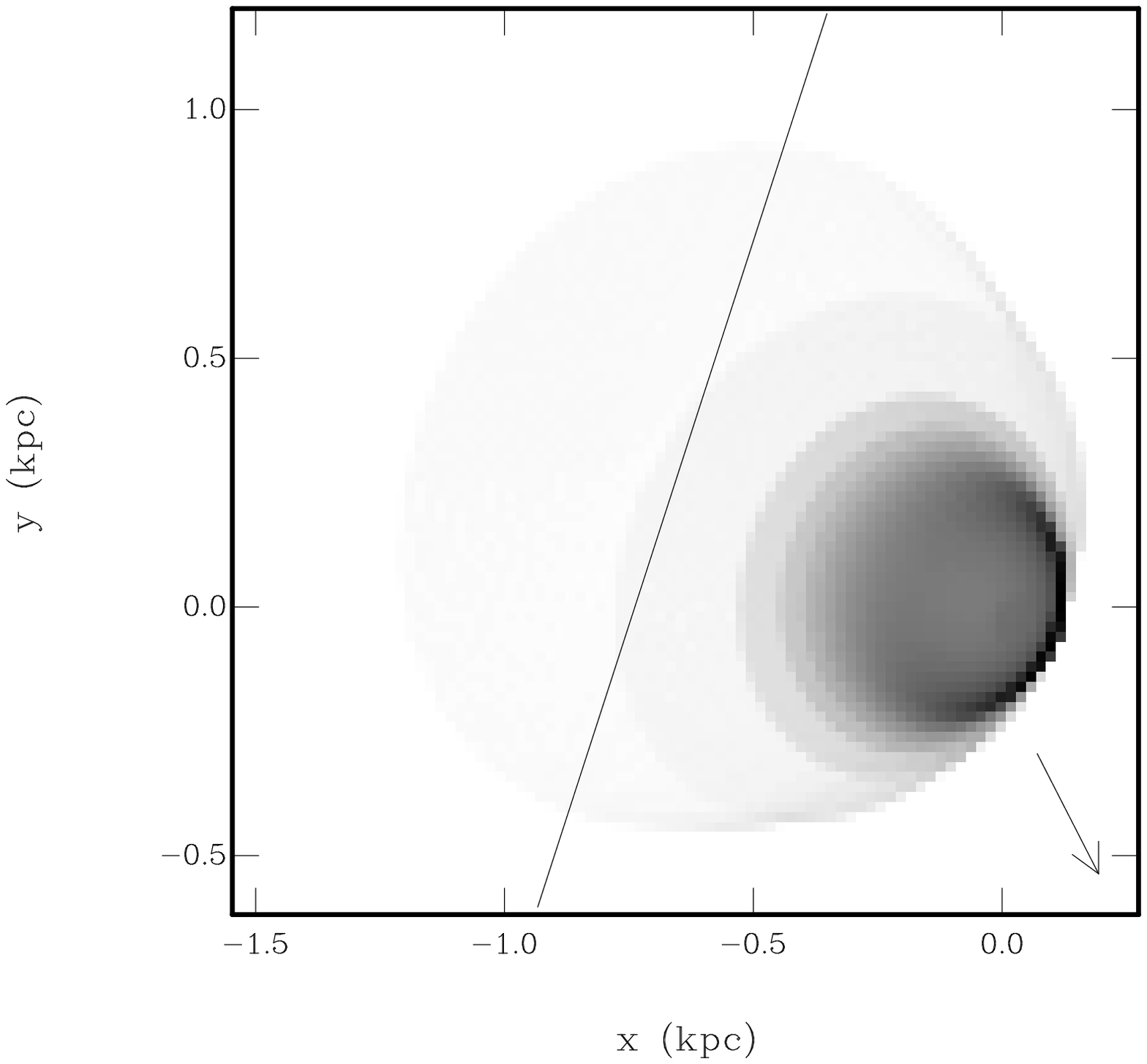, width=8cm, angle=0}
%\vskip-5cm
\caption[]{The projected column density of a shell expanding from 
$z = 100$ pc above
(left panel) or $z = -100$ pc below  (right panel) the symmetry plane of the 
galaxy.
North is up and West is right, the line shows the direction of the 
line of nodes, the 300 pc long arrow points towards the 
center of the LMC.}
\label{asym}
\end{figure*}

We try to reproduce not only the size of the 
shell at the right time and distances between fragments, but we 
would like to have  the correct orientation in the LMC and give
plausible arguments why the Sextant is only a fraction  of the total 
circle.
Here, projection effects play a role. In Fig. \ref{sym} we show 
the column density in the shell with the continuous input energy source 
of $10^{53}$ erg in the symmetry plane of the galaxy at t = 11 Myr, 
$\rho_{ext}(z=0) = 20$ cm$^{-3}$. The galaxy plane is observed in such a 
position that the line of nodes $\theta_0 = 162^0$ and the inclination 
$i = 33^0$. The column density is distributed in an elliptical ring. Its
ellipticity corresponds to projected dimensions of the 3D egg-shaped
shell (see R-z cuts in Fig. \ref{fig3}), not to the cosine of the 
inclination angle as would be the case of the projection of the in-plane
ring. There are 
two maxima in the column density resulting from projection effects. 
They correspond to  those directions where the path along the line of 
sight in 
the shell is long. Minima appear where the line of sight intersects the 
shell almost perpendicularly.

To get one-sided structures with one maximum in the column density, we 
calculated models with sources at different distances above or below the 
galactic plane. In Fig. \ref{asym} we show two cases with a 
continuous energy source of $10^{53}$ erg placed at $z = 100$ pc and $z = -100$
pc, $\rho_{ext}(z=0) = 20$ cm$^{-3}$. The expanding structure is asymmetrical 
because of  the asymmetry in the density distribution and
the gravitational potential. Its most dense part 
is at low $|z|$ distances near the symmetry plane and its 
low density parts open to high $|z|$ distances. The projected 
column density shown
in Fig. \ref{asym} has one maximum and it has the shape of an arc.
However, the orientation and opening of the arc is in both cases (with
the source above or below the symmetry plane) different from the observed 
position of the Sextant.

%% fig7
%%%%%%%%%%%%%%%%%%%%%%%%%%%%%%%%%%%%%%%%%%%%%%%%%%%%%%%%%%%%%%%%%%
\begin{figure}[htbp]
\leavevmode
\epsfig{file=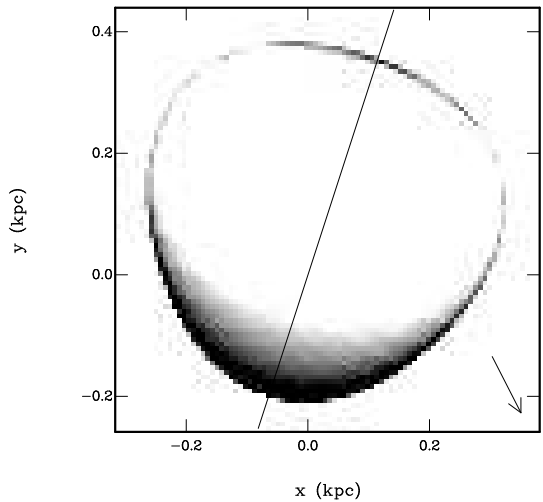, width=8cm, angle=0}
%\vskip-5cm
\caption[]{The projected column density of a shell expanding from an off-center 
position  inside  a preexisting  HI giant cloud.
North is up and West is right, the line shows the direction of the 
line of nodes, the 100 pc long arrow points towards the 
center of the LMC.}
\label{cloud}
\end{figure}

\begin{figure}[ptbh]
\leavevmode
\epsfig{file=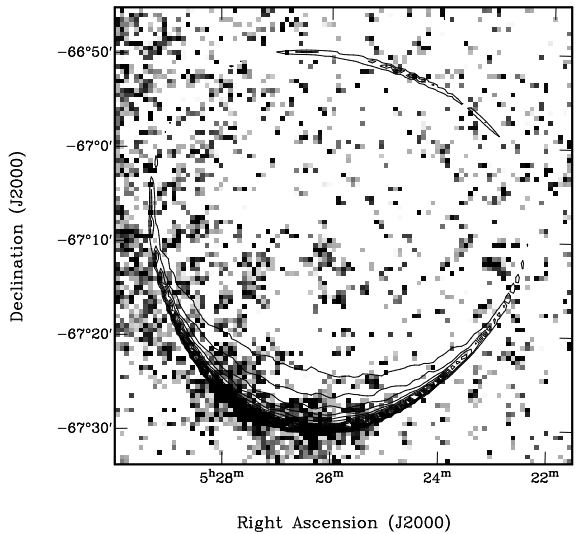, width=8cm, angle=0}
\caption[]{A part of the photographic image of the Constellation III region
including the Sextant arc. The original photograph from Boyden Observatory,
courtesy of Harlow Shapley, Harvard College Observatory,
has been put to computer readable form by Knut Olsen, Paul Hodge and Don
Brownlee. The photographic image is overlayed with 
isodensity lines giving the model column density (the maximum projected 
column density 
is 4.2 $\times $ 10$^{20}$ cm$^{-2}$, lines show 90\% , 80\% ... 10\% 
of this value).}
\label{sex}
\end{figure}

At the edge  of LMC 4 the HI density may have been increased by 
sweeping of the mass from the HI void as observed by Kim et al. (1998).
The HI collected there  probably fragmented and formed giant
clouds of increased HI density. We assume that the source of energy resided 
inside one of such HI clouds. In Fig. \ref{cloud}
we show results of a simulation where the shell formed around a continuous 
energy source.
The expansion started in  a
homogeneous spherical cloud of density 20 cm$^{-3}$ and radius 150 pc, where
the energy source has been placed   
at an off-center position (about 50 pc from the cloud center). 
The resulting projected column density (Fig.\ref{cloud}) 
does not depend strongly on the position of the source inside 
the parent cloud. The energy sources inside a rather large 
area NW from the giant cloud center produce the shape 
and orientation in the LMC similar to the Sextant. 
This is shown in Fig. \ref{sex}, where the model projected column density is 
plotted over the photographic image of the Sextant area.  

%Fig. 8
%%%%%%%%%%%%%%%%%%%%%%%%%%%%%%%%%%%%%%%%%%%%%%%%%%%%%%%%%%%%%%%%%

We conclude that 
shell forming a fraction of the complete circle in the correct 
position can be created by expansion which takes place in the 
NW part of a preexisting giant HI cloud at the edge of LMC 4.    

\section{Can GRB trigger and support star formation in galaxies?}

We have demonstrated that
the abrupt energy inputs related to GRBs form expanding shells, which can 
fragment and trigger star formation. An inspection of Table \ref{fragtab} 
shows that with energy  $E_0 = 10^{54}$ erg shells fragment
in any tested density of the ambient medium: $\rho_{ext} = 1 - 30$ cm$^{-3}$.
In our opinion, GRB are the likely cause of the supershell and the   
triggering of star formation, particularly when:
\begin{enumerate}
\item there is no stellar cluster inside a supershell or inside an arc of 
young OB stars (providing the normal IMF),
\item there is no evidence of HVC and/or neighbouring
galaxies,
\item there is no hot gas inside the shell, even in the case of a 
young supershell,
\item there is a system of multiple stellar arcs.
\end{enumerate}

The hypothesis of Blitz et al. (1999), that  HVCs are
debris after formation of the Local group of galaxies, suggests 
large distances to HVCs, which imply for a HVC the typical  mass of
$3 \times 10^8 M_{\odot}$ of neutral gas   
and the typical size of $ 25$ kpc. In such a case the 
theory, that the largest superbubbles may be created by impacts of HVCs, 
seems to be less plausible.    

Many supershells in M31 (Brinks \& Bajaja 1986) and 
M33 (Deul \& den Hartog 1990)
have a circular form and there are no visible  OB associations inside them.
A number of supergiant shells without
visible clusters is known in isolated galaxies, where the probability 
of HVC infalls is low, for example galaxies NGC 3044 
and NGC 4631 (see references in Efremov et al. 1998 and Loeb \& Perna 1998). 

Recently, a special photometric search for clusters, which could be the 
energy source of the supershells in the irregular dwarf galaxy Holmberg II,
was carried out by Rhode et al. (1999). 
They found only 4 out of 50 supershells with  clusters of 
suitable ages, which could have contained enough OB stars  
to form the observed supershells, assuming the normal IMF.
The best examples of supershells with no clusters are the most energetic ones
found in the low density environment at the periphery of Ho II, where no star 
formation occurs and where the presence of massive clusters is almost 
excluded.
Another result given by Rhode et al. (1999) is that nearly  no HVCs exist 
around Ho II, which may eventually be responsible for the observed 
supershells.
Hypernovae are also not the likely energy source, since they reside near the
star forming regions  (Paczynski, 1998), 
and there is no star formation near the supershells
at the periphery of Ho II. 
The energy source could be a merging event of a compact binary system. 
Supernova explosions, leading to the formation of compact components, 
may give the system  
the high space velocity of $\ge  $100 km s$^{-1}$ (Blaauw 1961). 
Then, the GRB  occurs far from the formation site as 
demonstrated e.g. by Lipunov et al. (1997).

The Sextant arc is one in the system of three stellar arcs in the LMC 4
region (Hodge 1967, Efremov \& Elmegreen 1998a). 
If the GRB theory of the origin is correct, then the progenitor 
of the Sextant 
arc (and progenitors of the other arcs as well) probably belonged to the
massive 2 Gyr old cluster NGC 1978 located 0.5 - 1.0 kpc from its center.
NGC 1978 has an elongated shape which may be the result of merging of
two clusters, or of  disk-shocking. Such events are able to eject many
stars (including compact binaries) from the cluster.
Actually, there is concentration of X-ray binaries in the discussed 
region. Also, one probable relict of the merging event of the compact binary
system, the Soft Gamma Repeater SGR 0526-66 in SNR N49, 
is at a distance of 18' from the cluster NGC 1978 (Efremov 1998, 1999a, b).  
  
At least three systems of multiple stellar arcs
are known  in the LMC, NGC 6946 and M 83
(Hodge 1967; Efremov 1999a).  
The opening angle in most of these arcs is  few dozen
degrees.
The strictly circular form of these arcs implies an origin from
the central source. Regular distances between clusters
along an arc suggest their formation by the gravitational  instability
in the swept-up gas ring (Sextant and two arcs in M 83).

Many questions related to the origin of supershells and triggered
star formation in connection to  GRBs remain open: 
\begin{itemize}
\item Can isolated galaxies with  active star formation be those in which
there were recently many GRBs?
\item Can GRBs  support and revive star formation activity  together
with OB stars and SNe?
\item Can GRBs be the initial source to  the star formation and turbulent
motions in gas disks of galaxies?
\end{itemize}
The last item (Hodge 1998, private communication) is quite attractive: 
compact binaries with
neutron stars may form in the process of dynamical evolution of
massive and old star clusters. The two compact components merge
and produce GRBs long after they escaped from the cluster. 
Those, which explode
within the gas disk, might trigger the turbulent motions and star formation
there.

Considering the large uncertainties in the theory of GRBs,
properties of supershells and stellar arcs suggested to be created
by GRBs may give essential constraints to the GRB theory.
The central angle of stellar arcs, for example, may be connected
somehow  with the beam angle of the GRB -
especially if they cannot be explained as projected shells 
created by sources 
outside the galaxy plane or by sources inside giant interstellar clouds.

Our simulations show one possible obstacle to GRBs playing an active role 
in 
star formation events in galaxies. The energy needed to create gravitationally
unstable shells by an abrupt explosion must be       very high 
($E_0 \ge 10^{53}$ erg; for low ISM densities even $E_0 \ge 10^{54}$ erg). 
There were indeed GRBs with energy $10^{53}$ and $10^{54}$ erg, such as 
GRB 971214 and GRB 990123 (Kulkarni et al. 1999), but the beaming of the
relativistic jets would imply that the emitted energy would be 
proportionally smaller. So, if the 
energy loss during the time interval between the burst itself and the 
time when the expanding shock forms, is large, or if the energy released 
by the burst is small, then the effect of GRBs on the host galaxy 
is minor and it can not induce  star formation.  

\section{Conclusions} 

We  discussed properties of HI shells connected either to GRBs or  OB 
associations: in the first case, there is an initial inflation or 
rapid growth of the shell followed by a slow expansion. 
In the second case the initial expansion is much slower, however, 
the expansion velocity decelerates to low values longer.
Supershells of diameters
larger than 500 pc are almost static if related to GRBs. 
If a supershell larger than 500 pc expands with a  velocity higher than 
$\sim $ 10 km s$^{-1}$, it is
undoubtedly related to an OB association (= continuous energy input). 

Using computer simulations of expanding shells in galaxies, we showed that
both the abrupt energy input connected to a GRB or a continuous energy input
connected to an OB association can trigger  star formation. 
Shells related 
to GRBs need the higher energy to
fragment into protoclusters than  shells connected to OB associations, 
where the lower 
energy is sufficient. The star formation 
triggered by GRBs can initiate or complement the star formation triggered
by OB associations.

As an example, we discussed the Sextant region in the vicinity of LMC 4. 
This arc of OB associations can be interpreted as triggered star formation
within the HI shell.   We found that both abrupt and
continuous energy inputs can correctly reproduce the radius, the age of the
arc and the average distance between OB associations. 

Partial shells formed by an energy source above or below 
the galactic plane are projected to the tangential plane perpendicular 
to the line of sight as arcs with apexes turned to or from 
the line of nodes of the LMC. 
This is not the case with the arcs in the LMC 4 region.
To reproduce the orientation and position of the Sextant we
have to assume, that the expansion took place inside a giant HI supercloud,
which had been previously formed at the rim of LMC 4.    

Clusters HS287 and HS288 are not in the very
center of the Sextant arc.
This may be explained as a consequence of the asymmetry of the shell: 
the projected structure (or its densest parts) 
has a center at a position different from the energy source. 

A comparatively large number of SNe is needed to fit the observed arc.
A cluster
producing 10 - 100 SN requires, with the normal IMF, many low-mass stars
in the same cluster. 
Unfortunately, there is no photometry for these clusters available to 
check their mass function. Because of the proximity of the massive star 
cluster NGC 1978, there is a high probability that
both compact binaries and hypernovae exist in this region (e.g. 
Chevalier \& Li 1999).  The occurrence of 
very massive clusters near the arc systems in LMC and NGC 6946 
shows that their
progenitors may be stellar objects.     

In any case, the presence of two or three more arcs composed of OB stars 
and clusters in the same
region near LMC 4 (Efremov \& Elmegreen 1998a), and the absence of similar 
arcs elsewhere in the LMC, can be better explained by a number of 
progenitors escaping from a single massive cluster, 
presumably NGC 1978. The recent discovery of the most massive young cluster 
in the galaxy NGC 6946, which is located near the only multiple arc system 
in this galaxy (Hodge 1967; Larsen \& Richtler
1999), supports the idea of the connection between multiple, 
not concentric arcs with progenitors escaping from one common massive cluster.

%\newpage
\begin{acknowledgements}
We would like to thank Bruce Elmegreen and Guillermo Tenorio-Tagle 
for careful reading
of the paper and the anonymous referee for valuable comments.
We are grateful to Sungeun Kim for the full text of her PhD 
thesis devoted to the high-resolution HI observation of the LMC and for 
providing us with the HI total column density map of the LMC.
Yu. N. E. appreciates the partial support by
the Russian Foundation of Fundamental Researches and the Council for the Support of Scientific Schools. He is also sincerely grateful to Jan Palou\v s and
to the staff of the Astronomical Institute of the Academy of Sciences of the 
Czech Republic for warm hospitality during his stay in Prague in
September 1998.
The authors gratefully acknowledge financial support by the Grant
Agency of the Czech Republic under the grant No. 205/97/0699 and
support by the grant project of the Academy of Sciences of the
Czech Republic No. K1-003-601/4.
\end{acknowledgements}

%\vskip1cm

{}

\end{document}